# Breve estado del arte en minería de información social:

Aplicación práctica en análisis de tendencias en legislativas francesas 2024. (w/source code)

José Alberto García Gutiérrez, *josea.garcia.gutierrez.edu@juntadeandalucia.es*

1. Introducción

El análisis de información social en redes sociales ha experimentado una evolución significativa en la última década impulsado por los avances en inteligencia artificial (en adelante IA) y machine learning (en adelante ML). Las redes sociales y las plataformas de microblogging han transformado la forma en la que las personas se informan y también las referencias que estos tienen a la hora de conformar sus propias opiniones, y las maneras en que estas se difunden hasta dar lugar a la creación de consensos en la sociedad moderna (Lazer et al., 2018; Martínez-Rojas et al., 2018). Hemos visto prueba de esto durante los procesos electorales donde gobiernos, partidos políticos, y agentes sociales se han lanzado a la propagación interesada de información sesgada o manipulada cuidadosamente a la vez que, en otras ocasiones, estas redes han sido el único medio de difusión de noticias que eran muy matizadas o directamente silenciadas por las líneas editoriales de los medios mas convencionales. Ambas situaciones, igualmente anómalas, denotan la importancia de estas redes a la vez que las sitúan como poderosas herramientas que se han convertido en autenticas armas de doble filo en la era digital (Gerbau & Fourner, 2018). La ingeniería de la reputación y el conocimiento profundo de cómo se propagan las opiniones en las redes sociales también resulta útil en situaciones de crisis o emergencia, donde el control de la desinformación y la transmisión eficaz de instrucciones o guías a la población puede ayudar a hacer mas efectivas las medidas gubernamentales o mejorar la respuesta y las tasas de supervivencia. Shahbazi et al., tratan este tema en diferentes estudios en 2013, 2023 y 2024 abordando en detalle lo ocurrido en experiencias reales como los casos de la pandemia covid, los terremotos de Irán e Iraq de 2019 y la última crisis financiera. Estas cualidades para llegar a las masas en detrimento de otras medios de comunicación mas tradicionales también han hecho que gobiernos de todo tipo inicien planes para intervenir en mayor medida las interacciones sociales y que medios como Twitter (X.com) estén implicadas con frecuencia en distintas campañas de desinformación o manipulación, ya que ofrecen un terreno fértil para la difusión de narrativas falsas o distorsionadas, que en ocasiones se utilizan para blanquear la gestión de un determinado personaje u organismo, o potenciar aspectos positivos sobre la memoria colectiva de los individuos. Las teorías de la conspiración y la reinterpretación sesgada de eventos pasados encuentran en estas plataformas un público amplio y poco crítico, lo que puede llevar a una narrativa errónea de la historia o, aún peor, a la utilización de estas redes para el adoctrinamiento ideológico, para influir en las aulas o para reforzar o hundir la reputación de determinados personajes o personalidades públicas, enardeciéndolos, ridiculizándolos, o directamente cancelándolos. Esto es especialmente preocupante en contextos donde se trata de imponer una versión revisada o manipulada de ciertos acontecimientos para así repercutir en la reputación de ciertos actores o construir una identidad cultural o política en una determinada comunidad mas afín a ciertos intereses (Vosoughi, Roy y Aral, 2018).

Antes de 2013, el análisis de información social se basaba principalmente en métodos manuales y artesanales. Las técnicas de análisis de sentimientos tempranas utilizaban listas de palabras positivas y negativas y anotación sintáctica, lo que las dotaba de una capacidad limitada para comprender contextos, ironías o lenguaje informal o especializado (Pang & Lee, 2008). Aunque algo más complejos, los sistemas de reputación de la época se centraban en el conteo y la ocurrencia de ciertos términos en comentarios positivos y negativos, o el seguimiento que se hacia de distintos términos, sin profundizar en el contenido semántico o el impacto a largo plazo de las opiniones (Barberá & Martinez, 2015). Estas limitaciones dificultaban la gestión proactiva de la reputación, relegándola a una respuesta reactiva ante crisis ya evidentes.

La llegada de plataformas como Twitter (X.com) y Facebook (Meta), junto con el aumento exponencial de datos, impulsó la necesidad de métodos más sofisticados.

2. Estado del arte en análisis de datos provenientes de redes sociales.

A partir de 2012-13, la IA y el desarrollo que sufrió el ML con la implementación de grandes centros de computación como el *Microsoft Data Center Dublin*, o en España el *Barcelona Supercomputing Center* (BSC) y la masificación de las fuentes abiertas de datos transformaron el análisis de la reputación en redes sociales, impulsado significativamente por los avances en el procesamiento del lenguaje natural (NLP). La publicación de los primeros modelos de *embedding* como Word2Vec en el año 2013 fue sin duda un hito importante y representó un avance significativo en el análisis de los datos sociales y de redes como Twitter al permitir crear representaciones vectoriales de los textos conservando una gran parte de su semántica que ahora se aprendía por sí sola desde grandes volúmenes de textos sin etiquetar. A partir de este punto, el avance fue exponencial. A Word2Vec le siguió en 2014 el modelo de embedding de Stanford (GloVe), que era capaz de capturar relaciones globales y locales entre palabras. También de ese año son los primeros modelos Seq2Seq, que aprovechaban la arquitectura codificador-descodificador de (Sutskever et al., 2014) para hacer correspondencias entre textos codificados desde diferentes dominios, mejorando enormemente, entre otras cosas, los modelos de traducción automática y permitiendo entrenar con *datasets* multilenguaje sin tener en cuenta la barrera del idioma.

En lo que respecta al análisis de tendencias de opinión, de estos años hay trabajos interesantes, entre los que podríamos destacar el trabajo de (Spina et al., 2014), que introdujo técnicas de agrupamiento según el contenido y la estructura de los tweets utilizando técnicas de embedding y funciones de similitud basadas en características textuales y contextuales de los documentos y no el enfoque tradicional basado en el modelo de asignación de Dirichlet latente (LDA) para la clasificación automática de grandes conjuntos de documentos propuesto por (Blei et al., 2003). Este enfoque mejoró la detección de temas relevantes y la gestión de la información, facilitando la identificación de patrones y tendencias en la opinión pública. Y también el trabajo de (Peetz et al., 2015), que reconocieron que la percepción de la reputación puede cambiar con el tiempo, por lo que propusieron modelos que consideraban la evolución de la polaridad de los tweets a lo largo del tiempo. Esto permitió identificar tendencias y patrones en la opinión pública que no se podían detectar con modelos estáticos. También incorporaron la influencia de las respuestas de los usuarios a los tweets originales en sus modelos de clasificación. Esto permitió comprender mejor cómo las interacciones entre usuarios amplificaban o atenuaban la polaridad de los mensajes originales, ofreciendo una imagen más completa del impacto de las interacciones en línea.

El siguiente salto se produjo a finales de 2015 con la presentación del modelo *Transformer*, una arquitectura que permitía trascender la codificación basada en palabras y aprender a reconocer textos como gramáticas de contexto libre formadas por secuencias de tokens, posibilitando aprender representaciones contextuales de texto como secuencias de tokens. Utiliza mecanismos de atención que capturan dependencias a largo plazo en el texto. Los llamados mecanismos de atención están en la base de todas las arquitecturas modernas y, aunque son uno de los mecanismos fundamentales en el funcionamiento de los Transformer, cobraron especial popularidad a raíz del paper "*Attention Is All You Need*" de (Vaswani et al., 2017), sentando las ideas que han permitido a los investigadores desarrollar grandes modelos de NLP más precisos y eficientes. Al introducir un mecanismo de atención basado en el modelo de codificador-decodificador, los autores demostraron que era posible entrenar modelos de NLP que podían descubrir por sí mismos, usando mecanismos fáciles de implementar, las partes más relevantes del texto de entrada. Esto condujo a una mejora significativa en el rendimiento de una variedad de tareas de NLP, incluyendo el reconocimiento de entidades, tareas de inferencia, equivalencia semántica y la reformulación, resumen e indexado de textos, procesamiento de textos desestructurados y las tareas de respuesta a preguntas, todas áreas de interés en el análisis de información social.

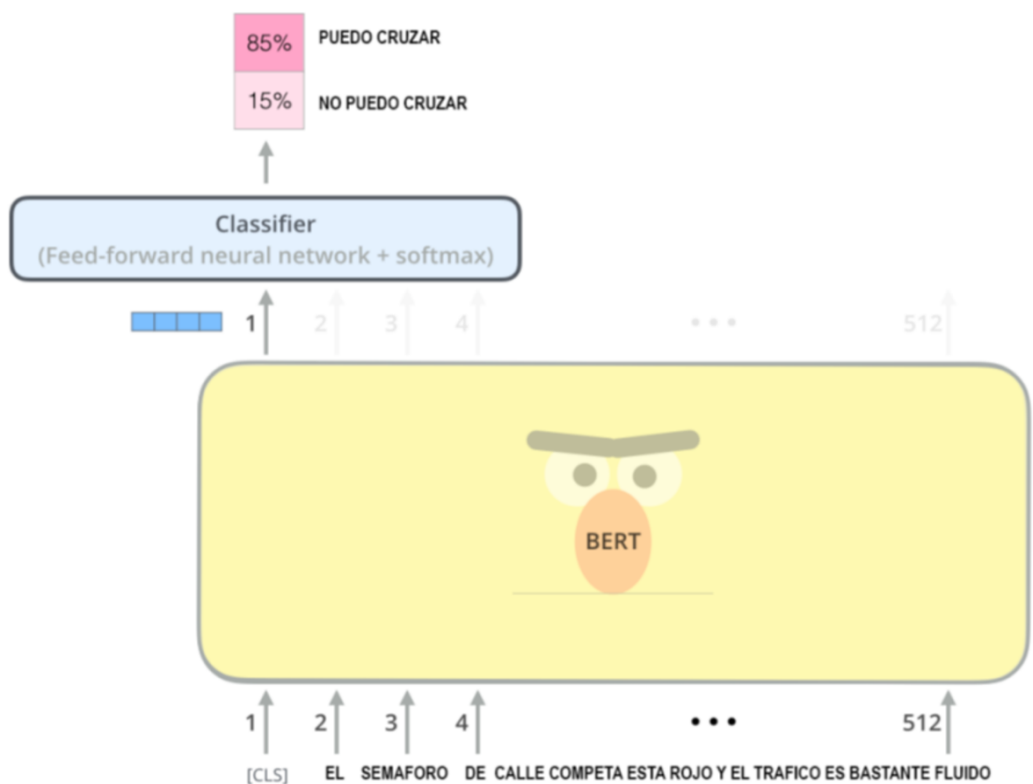

*Ilustración 1. Ejemplo de arquitectura con codificador BERT*

Más tarde, en 2018, el modelo BERT de Google se presentaría como el primer modelo con entrenamiento bidireccional (permitiendo romper las relaciones de linealidad de los textos) y arquitectura reutilizaban preentrenada. Lo cual daba a los investigadores un poderoso codificador universal para textos que podían aplicar a todo tipo de usos finales con un pequeño ajuste fino, integrando rápidamente el estándar en clasificación de textos, análisis del sentimiento, detección de emociones y clasificación de temática *zero-shot*. Estos modelos han mejorado la capacidad de las máquinas para comprender el significado del texto en las redes sociales, lo que ha permitido un análisis más profundo de la opinión pública y la reputación en línea.

3. Áreas activas de investigación

3.1 Modelos de Embedding Complejos:

Como hemos visto en apartados anteriores, en los últimos 10 años la introducción de modelos de embedding complejos como Word2Vec y GloVe ha acaparado gran parte del protagonismo con relación al análisis de reputación en redes sociales. Estos modelos permiten representar palabras y frases como vectores numéricos, capturando relaciones semánticas y contextuales que antes eran difíciles de codificar (Mikolov et al., 2013; Pennington et al., 2014).

El uso de métodos de incrustación (*embedding*) tiene algunas ventajas importantes. Los modelos de embedding permiten a las herramientas de análisis de reputación comprender mejor el significado sutil del texto, incluyendo ironía, sarcasmo y jerga (Liu, 2021). Los embedding también permiten un análisis más preciso: Poder extraer una mayor carga semántica de los textos permite un análisis mucho mas profundo y esto conduce a un análisis de sentimientos y una clasificación de temas más precisos, lo que proporciona información más confiable sobre la reputación online (Devlin et al., 2018). Y facilitan la detección de temas emergentes y tendencias en las redes sociales, contribuyendo y facilitando el análisis reputacional y el seguimiento de tendencias y permitiendo a las empresas identificar problemas potenciales de reputación de manera proactiva (Rashid et al., 2024).

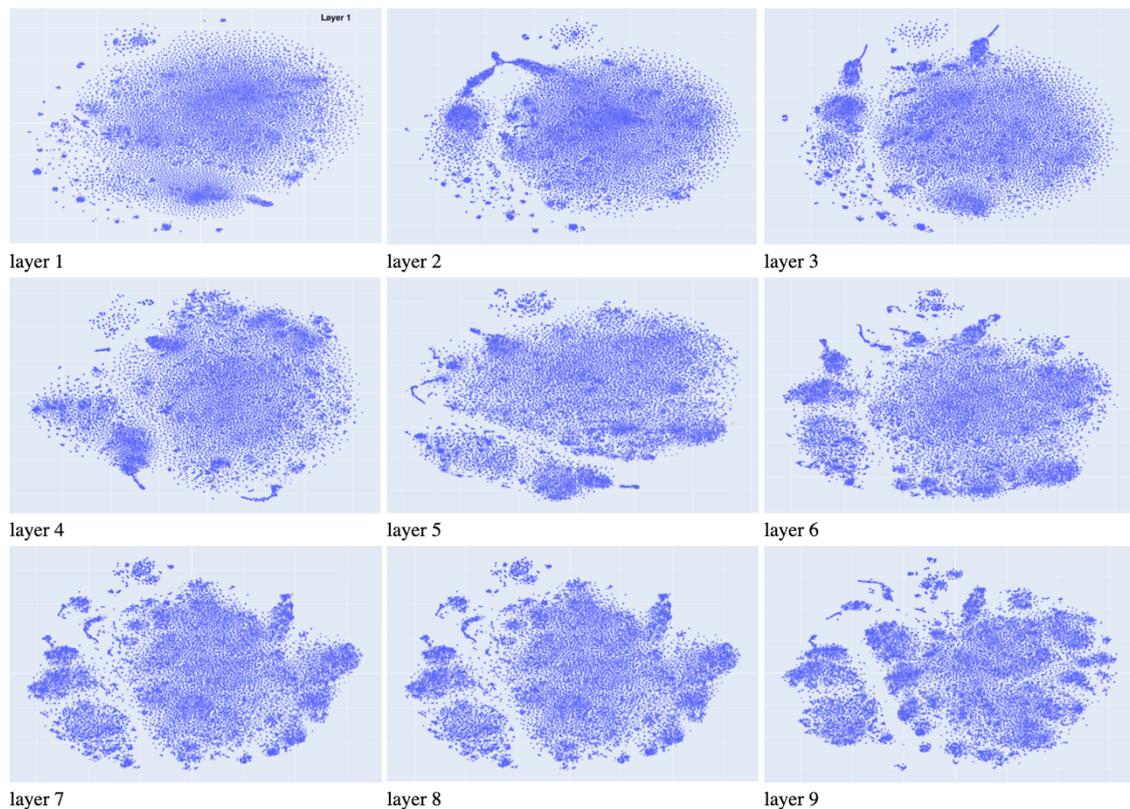

*Ilustración 2. Proyección usando TSNE del espacio de embedding de cada una de las 12 capas Transformer de la arquitectura BERT (Devlin et al., 2018).*

Algunos retos y cuestiones de investigación abiertas hoy en día con relación a los embedding incluyen:

- Adaptación a diferentes dominios: Los *embeddings* generales pueden no ser óptimos para todos los dominios. La investigación se centra en crear *embeddings* específicos para dominios (finanzas, medicina, etc.) que capturen mejor las particularidades de cada área.

- Multimodalidad: Con el surgimiento de modelos multimodales y en un contexto como las redes sociales de naturaleza claramente multimodal y multimedia, integrar texto con otros tipos de datos (imágenes, audio) para mejorar la comprensión del contexto general es un área muy activa sobre la que se sigue trabajando.

- Soporte Multidioma y Embeddings multilingües: Mejorar la representación y comprensión de múltiples idiomas en un solo espacio de *embeddings*, facilitando la traducción automática y la transferencia de conocimiento entre lenguajes.

- Lograr cada vez mayor precisión en el análisis de sentimiento y emociones: *embeddings* más precisos pueden captar matices emocionales, mejorando la detección de sentimientos en las redes sociales.

3.2 Bases de Datos Vectoriales:

La llegada de los grandes modelos de lenguaje (LLMs) como BERT y posteriores, ha permitido crear representaciones vectoriales de alta dimensión para palabras, frases y documentos, capturando relaciones semánticas y contextuales con cada vez mas fidelidad. Esto facilita la creación de bases de datos vectoriales que permiten búsquedas más precisas y relevantes en grandes volúmenes de texto (Devlin et al., 2018). Las bases de datos vectoriales almacenan y recuperan documentos en función de su similitud semántica, en lugar de simplemente por coincidencia de palabras clave. Esto es crucial para tareas como la búsqueda de información y la recuperación de documentos relevantes (p.e Faiss, una biblioteca de búsqueda de similitud desarrollada por Meta AI). Aplicados al análisis de información social, los LLMs permiten identificar patrones y tendencias en series cruzadas y grandes volúmenes de datos desestructurados como los provenientes de redes sociales, identificando temas emergentes más rápidamente y con mayor precisión (Liu et al., 2019). Así como también permiten una mejor clasificación de sentimientos y emociones en los textos, capturando matices que los enfoques anteriores no podían detectar. Esto es fundamental para el análisis de la reputación en las redes sociales.

Las bases de datos vectoriales y los LLMs permiten analizar cómo se propagan los mensajes y cómo las interacciones entre usuarios influyen en la reputación. Como ejemplo, (Zhang et al., 2020) utilizaron *embeddings* generados por LLMs para analizar la propagación de información en Twitter). Al incorporar información de estas bases de datos, las herramientas de análisis de reputación pueden enriquecer su comprensión del contexto en el que se utilizan las palabras y frases, mejorando aún más la precisión del análisis (Färber et al., 2022). Las bases de datos vectoriales también facilitan la identificación de entidades nombradas, como personas, lugares y organizaciones, en las redes sociales, lo que permite un análisis más granular de la reputación y de como esta se ve influenciada por las distintas tendencias y corrientes de opinión a lo largo del tiempo.

2.3 Grandes Modelos de Lenguaje (LLM):

Desde el lanzamiento de BERT en 2018, los últimos 6 años han venido marcados por la aparición de mas y mejores modelos del lenguaje. Tras la aparición de BERT y sus muchas variantes (AlBERT, RoBERT, Distillbert o XLNet) siguió la llegada de GPT y poco tiempo después de GPT-2 (2019), y ya con GPT-3 (2020) el tamaño del modelo había crecido hasta los increíbles 175K millones de parámetros. Cada uno de estos modelos ampliaba sus capacidades en cuanto a precisión, calidad de conocimiento extraído, tiempos de entrenamiento e inferencia, coste computación y tamaño de ventana de contexto. Mejorando sus mecanismos de atención que ahora abarcan atención multinivel, la forma de sus embedding utilizando distintos modelos de incrustación especializados en tareas concretas y optimizando el tamaño de las representaciones internas de los textos para obtener modelos más compactos y con menores tiempos de inferencia.

Aunque no está en las funciones de un LLM comprender el lenguaje como tal, su arquitectura basada en Transformers se especializa en aprender gramáticas, autodescubriendo sus elementos constitutivos, su semántica y sus relaciones a distintos niveles, lo que les hace dar respuesta a necesidades específicas y ser interrogados sobre tareas complejas y muy dirigidas. Esto unido a su enorme capacidad de computo y a que son entrenados con decenas de terabytes de ejemplos hace que sean capaces de aprender no solo la sintaxis y usos de un lenguaje sino sutilezas de los propios hablantes como ironías, hipérboles o metáforas, los que las hace ideales para el análisis de texto desestructurado proveniente de fuentes informales como X.com (Twitter).

La aparición de LLM como BERT, GPT y mas recientemente LLaMA y Gemini-LaMDA ha impulsado aún más el análisis de la reputación en redes sociales (Devlin et al., 2019; Brown et al., 2020; Akhtar et al., 2021). Los LLM poseen una capacidad sin precedentes para comprender el lenguaje natural, incluyendo matices contextuales, intenciones y emociones (Rogers et al., 2022). Esto permite un análisis más profundo de las conversaciones en redes sociales, identificando opiniones matizadas, detectando sesgos y comprendiendo las motivaciones detrás de los comentarios (Brundage et al., 2021). Los LLMs también pueden generar contenido textual, incluyendo informes, resúmenes de indicadores, o información destinada a cuadros de mando empresariales o estratégicos (Agnihotri et al., 2022).

Algunos retos que todavía tienen los LLM por delante y que aglutinan la investigación en el área incluyen:

- La tendencia de los modelos del lenguaje a generar respuestas plausibles, pero no fidedignas a preguntas de las que no poseen suficiente conocimiento. Este problema se ha tratado de solventar de diferentes maneras, entre ellas la incorporación en la respuesta de conocimiento semántico o ontológico. La validación de las respuestas mediante bases de datos especializadas. O la generación de grafos de conocimiento que permitan guiar la búsqueda del modelo. Por ejemplo, (Urena et al., 2020; Bathla et al., 2021) proponen modelos con un enfoque basado en grafos para el análisis de la reputación en redes sociales que considera la influencia del usuario. El modelo no solo considera el contenido de los tweets, sino también la red de relaciones entre usuarios para inferir la reputación de una entidad. Este enfoque permite comprender mejor cómo las opiniones se propagan en la red social y cómo influyen en la reputación general.

- Las redes sociales por naturaleza utilizan un enfoque multimodal y multilenguaje. Distintos tipos de lenguajes y vocabularios especializados se mezclan con expresiones coloquiales, lenguaje informal, términos prestados de distintos idiomas y fotografías, figuras, ilustraciones audios y vídeos de distintas características. Comprender esta modalidad y incorporarla a la representación interna de los grandes modelos permitirá comprender mejor la verdadera naturaleza de las interacciones en estos medios y hacer un análisis reputacional mucho mas preciso y realista (Das & Singh, 2023; Uppada & Patel, 2023). Recientemente, modelos como GPT 4-o parecen haber comprendido esta problemática y los nuevos modelos colocan la multimodalidad como una pieza clave para generar contenidos equiparables en calidad al análisis humano.

- En áreas como el análisis de reputación no nos basta con obtener un análisis superficial de la evolución o las tendencias. El objetivo de este análisis es la previsión, ser capaz de prever y reaccionar a como se van produciendo los flujos de información social y para ello será necesario obtener una explicación o una descripción causal de los cambios que se van observando. La AI explicable es uno de los mayores retos a los que se enfrentan los modelos del lenguaje que por su forma de entrenamiento a veces generan resultados que resultan sorprendentes y inesperados incluso para los expertos en el área (Jiang et al., 2023; Chen et al., 2023). Estos modelos no solo buscan predecir la reputación de un organismo o una entidad, sino que también explica las razones detrás de sus predicciones.

Este enfoque es crucial para comprender mejor los factores que influyen en la reputación y para tomar decisiones informadas sobre la gestión de esta.

4. Elaboración de un pequeño proyecto de minería de información social

Una vez hemos repasado el estado del arte del análisis de información social en redes, proponemos un pequeño ejercicio aplicado en el que empleamos las técnicas anteriormente tratadas para analizar el nivel de seguimiento en redes sociales de los programas electorales de los principales partidos políticos que concurren a las elecciones legislativas francesas del 30 de junio de 2024.

4.1 Objetivos del estudio

Si acudimos a servicios de estimación de tendencias como Google Trends podemos ver como el interés por el partido de Marine Le Pen parece mantenerse alto frente marcando una importante ventaja en cuanto a presencia en redes sociales a pesar de los últimos sondeos preelectorales con relación a la coalición de izquierdas liderada por Mélenchon y al partido gubernamental de Emmanuel Macron. (Carral & Tuñon, 2020) apuntan a una ventaja que alcanza incluso el 400% en redes sociales de alta difusión como Twitter (X.com) y a un notable incremento en el número de retweets y en la difusión de los mensajes de campaña de los partidos de extrema derecha y extrema izquierda frente a otros partidos con esquemas más tradicionales. La ilustración 3 (arriba) muestra como esta inercia se confirma si tomamos los datos facilitados por Google a través de su herramienta se seguimiento de tendencias. Y observamos también como, mientras el voto gubernamental se concentra en regiones como Île-de-France, el seguimiento del partido de Le Pen es más trasversal y se distribuye de forma mucho mas uniforme (ilustración 3, abajo).

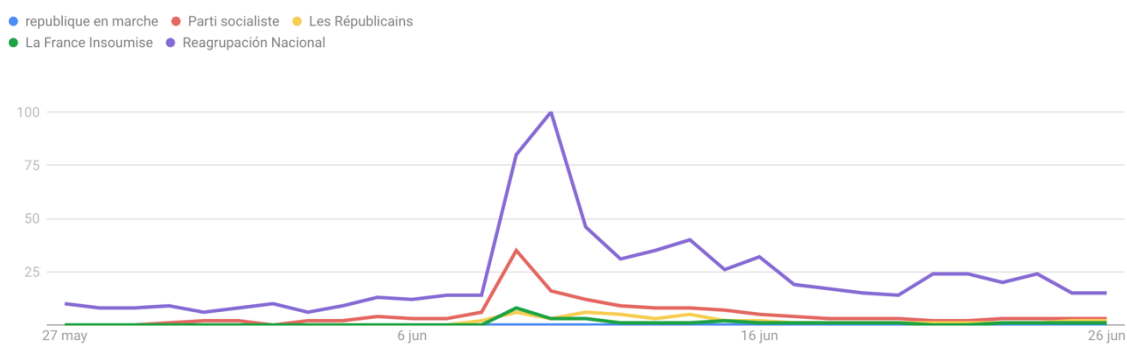

*Ilustración 3. Variación del interés en las diferentes candidaturas de la semana de campaña (arriba), y concentración de voto de En Marcha! (abajo derecha) y Reagrupación Nacional (abajo izquierda).*

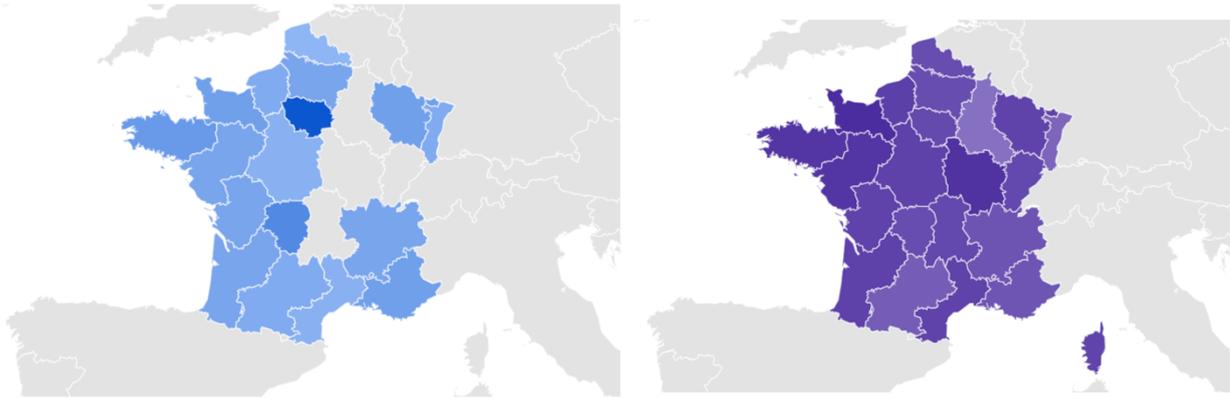

El objetivo del estudio es confirmar estas si estas conclusiones pueden extrapolarse a redes sociales menos masivas y con menos contaminación ideológica. Para ello, utilizando herramientas de procesamiento de lenguaje natural (PLN), investigamos el grado de aceptación de los mensajes políticos emitidos por estos partidos a través de los comentarios y reacciones de los propios usuarios a las noticias que van generando los medios durante la campaña. Nuestro enfoque se centra en identificar las dinámicas de interacción y respuesta de los usuarios frente a diversas propuestas políticas, permitiendo así una comprensión más profunda de las tendencias y preferencias electorales en el contexto digital

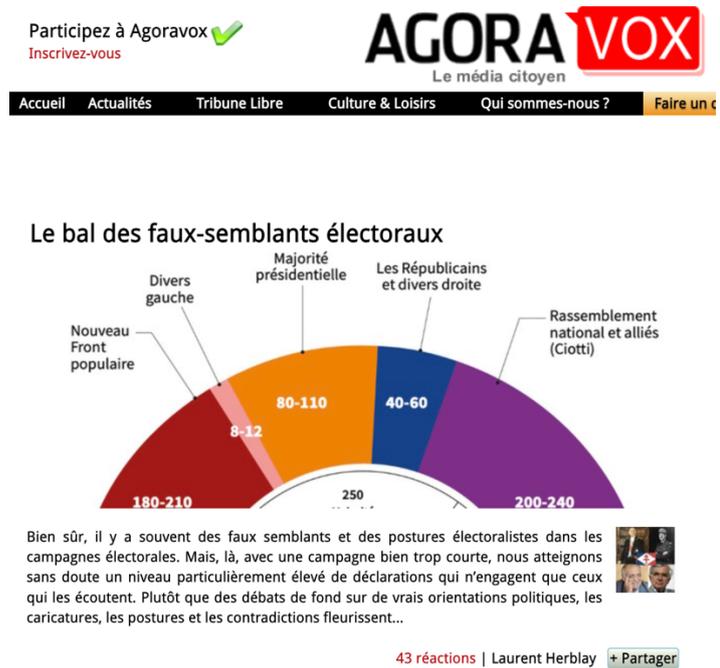

*Ilustración 4. Portada principal de la red AgoraVox.*

de internet. Para facilitar el trabajo utilizaremos una red social especializada en comentarios y noticias de índole política. *AgoraVox* es una red social donde los usuarios pueden compartir, comentar y elevar noticias, bastante similar a la red *Meneame* en España.

4.2 Materiales y métodos

Para realizar un análisis básico de las opiniones de los electores basándonos en las opiniones y reacciones que acompañan a las noticias que son tendencias en AgoraVox utilizaremos diversas librerías de scraping para acceder a las diferentes publicaciones en los distintos canales de noticias de AgoraVox y recuperar los comentarios y reacciones de los usuarios de la red a lo largo del tiempo. A continuación, utilizamos un LLM (gran modelo del lenguaje), en este caso la última versión de gpt para procesar los distintos programas electorales que los partidos políticos han presentado durante la campaña electoral y adaptaremos el modelo para que nos sirva para predecir la tendencia política de cada uno de los usuarios y sus reacciones. GPT-4, desarrollado por *OpenAI*, es un modelo de lenguaje autoregresivo que representa un salto significativo en escala y capacidad respecto a sus predecesores, utilizando una arquitectura Transformer con mas de 175 mil millones de parámetros (se desconoce su tamaño exacto, pero si se sabe que es mayor que gpt-3 que partía de esa base).

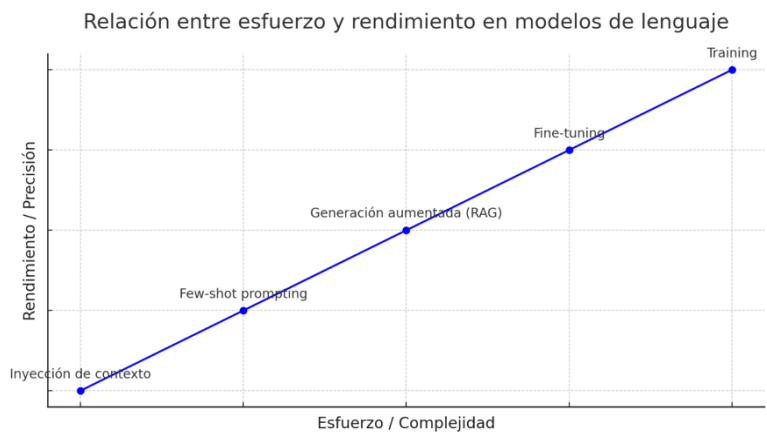

*Ilustración 5. Relación esfuerzo/rendimiento de los distintos tipos de ajuste fino sobre modelos LLM.*

A diferencia de BERT y otros modelos basados en *embeddings* contextuales, GPT es entrenado con un objetivo autoregresivo que predice el siguiente token en base a todos los anteriores, lo que le permite generar texto de manera continua. También al ser notablemente más grande y permitir contextos de entrada mucho mayores permite adaptarse mejor a tareas específicas. Los modelos GPT pueden realizar muchas tareas de procesamiento de lenguaje natural simplemente alterando las indicaciones del texto de entrada, sin entrenamiento específico para tareas individuales (conocido como "*zero-shot learning*"). Pero cuando queremos aportar al modelo conocimiento actualizado, limitados a áreas muy delimitadas o queremos que este indexe nuevo conocimiento a partir de documentos o bases de datos externas se hace necesario algún tipo de reentrenamiento ligero (*fine-tunning*). Cuando la terea que se requiere no supone el aprendizaje de nuevas reglas gramaticales ni aprender un nuevo lenguaje, generalmente es suficiente con la forma más barata de reentrenamiento ligero conocida como inyección de contexto (véase ilustración 5). En nuestro caso y al tratarse de una tarea que no es de generación de texto, complementamos la inyección de contexto con la técnica conocida como "*few-shot prompting*" lo que nos permitirá suministrar al modelo un gran número de ejemplos de clasificación para la tarea que le requerimos.

4.3 Recopilación de datos y creación del corpus de trabajo

Para construir nuestro corpus utilizamos el directorio RSS de AgoraVox para obtener los enlaces de las publicaciones que son *Trending topic* en la plataforma para las diferentes temáticas.

```
<item xml:lang="fr">
  <title>Manifeste pour la Nouvelle Gauche (Considérations militantes)</title>
  <link>https://www.agoravox.fr/actualites/politique/article/manifeste-pour-la-nouvelle-gauche-255441</link>
  <guid isPermaLink="true">https://www.agoravox.fr/actualites/politique/article/manifeste-pour-la-nouvelle-gauche-2554</guid>
  <dc:date>2024-06-26T16:48:16Z</dc:date>
  <dc:format>text/html</dc:format>
  <dc:language>fr</dc:language>
  <dc:creator>Alain Malcolm</dc:creator>
  <description>1. Dès les années 1970 il y eut ce que la force des choses convint de nommer « Nouvelle Droite » et qui par la force des choses (une simple concession au Zeitgeist de la part d'Alain de Benoist qui a été récupérée au balisage militant du terrain parfaitement moutonnier). 2. Quid de la Nouvelle Gauche ? Dans un premier temps on pour qui se nommèrent bien (...) - <a href="https://www.agoravox.fr/actualites/politique/" rel="directory">Politique</a></description>
  <enclosure url="https://www.agoravox.fr/IMG/jpg/creco-bretone-rouge-nouvelle-gauche-nouvelle-droite.jpg" length="573
</item>
<item xml:lang="fr">
  <title>Législatives 2024 (10) : il était une fois Jordan Bardella, Gabriel Attal et Manuel Bompard</title>
  <link>https://www.agoravox.fr/actualites/politique/article/legislatives-2024-10-il-etait-une-255436</link>
  <guid isPermaLink="true">https://www.agoravox.fr/actualites/politique/article/legislatives-2024-10-il-etait-une-2554</guid>
  <dc:date>2024-06-26T16:10:26Z</dc:date>
  <dc:format>text/html</dc:format>
  <dc:language>fr</dc:language>
  <dc:creator>Sylvain Rakotoarison</dc:creator>
  <description>« Je propose (…) une règle d'or : pas d'augmentation d'impôts pour les Français. C'est n'y aura pas, si les candidats Ensemble pour la République forment une majorité, d'augmentation des impôts de années (…), on a supprimé la taxe d'habitation, compensée aux collectivités locales, supprimé la redevance audio auparavant, vous avez tous les (...) - <a href="https://www.agoravox.fr/actualites/politique/" rel="directory">Polit</a></description>
```

*Ilustración 6. Fragmento de uno de los canales RSS para la categoría de política francesa.*

A partir de estas URL accedemos a cada una de las publicaciones y mediante web scraping utilizando las librerías *Feedparser*, *Request*, *BeutifulSoup* y NLTK parseamos el documento y extraemos los comentarios junto con los metadatos asociados a la entrada.

**Eric F** 26 juin 14:23 ★★☆ (3 votes)

@Fergus
il est naturel que les instituts de sondage fassent leur job et proposent des projections. Vous noterez que l'hypothèse haute de la gauche et l'hypothèse basse des souverainistes se croisent, rein n'est absolument joué.

Macron pense que la gauche éclatera, et qu'il se ralliera la droite de la gauche (et les zécolos) et la gauche de la droite (tout est relatif).
Il aura aussi la possibilité de prolonger le gouvernement en place "pour expédier les affaires courantes" comme ça s'est passé (assez bien) en Belgique pendant deux ans, ou désigner un gouvernement de techniciens (genre Trichet ou Breton) ;
Au pire du pire (complotistons un peu pour rigoler) il pourra acter de impossibilité d'une majorité et activer l'article 16 (pouvoirs spéciaux).

Ah quel suspense, et que c'est beau le pluralisme qui n'existe nulle part ailleurs dans le monde....

Répondre  Signaler un abus  Lien permanent

**Fergus** 26 juin 16:33 ★★☆ (2 votes)

Bonjour, Eric F

Ce scénario est possible, mais ne pourrait voir le jour que si le pays se trouve bloqué durant des mois par des motions de censure — et donc des chutes de gouvernement — à répétition.

Je ne crois pas trop à un scénario à la belge : ce n'est pas dans nos mentalités.

Quant à l'article 16, Macron ne pourrait pas l'activer en l'absence de motifs suffisamment « *graves* ». Et la durée de sa mise en oeuvre serait de toute façon trop limitée pour attendre la présidentielle de 2027. Au mieux de nouvelles législatives en 2025.

*Ilustración 7. Ejemplo de comentarios y votaciones de usuarios en la plataforma.*

Por limitaciones de tiempo y recursos los datos fueron recopilados entre los días 24 de junio y el 27 de junio de 2024 y no re realizó una búsqueda histórica fuera de este rango de fechas.

```
21 ▶        <COMENTARIO> ··· </COMENTARIO>
27 ▶        <COMENTARIO> ··· </COMENTARIO>
33 ▶        <COMENTARIO> ··· </COMENTARIO>
39 ▶        <COMENTARIO> ··· </COMENTARIO>
45      </COMENTARIOS>
46   </NOTICIA>
47 ▼ <NOTICIA>
48      <TITULO>Macron: Il a vendu la France aux Américains</TITULO>
49      <CATEGORIA>Política francesa</CATEGORIA>
50      <FECHA>2024-06-20</FECHA>
51 ▼    <RESUMEN>"Il a vendu la France aux Américains" : Révélations s
           (...)
52              réseaux. De la Silicon Valley à l'Élysée en passant
53              voici le récit peu connu des réseaux américains qui
                carrière
54              politique d'Emmanuel Macron. "Macron l'Américain", l
                documentaire
55              de Off Investigation retrace l'histoire d'une fascin
                d'un
56              président pour les Etats-Unis et ses géants du numér
57              surtout comment le Président de la République a déro
                rouge
58              aux GAFAM en démolissant méthodiquement le droit du
59              A travers ça le documentaire démontre plus largement
```
*Ilustración 8. Resultados del proceso de scraping en formato XML antes del preprocesado.*

### 4.4 Limpieza y anotación de los datos

A continuación, los datos fueron sometidos a varios procesos de limpieza para eliminar cualquier posible código HTML, referencias a recursos externos, código javascript, información de estilos, etc. Y el texto plano de los comentarios se tokeniza y se eliminan aquellos tokens que corresponden a texto superfluo (stopwords), que no alcanzan un umbral mínimo de cobertura en los textos, o todo lo contrario, que aparecen con demasiada frecuencia.

```python
# Funciones de limpieza de entradas
def limpiar_texto(texto):
    stop_words = set(stopwords.words('french'))
    palabras = word_tokenize(texto)
    palabras_limpias = [palabra for palabra en palabras if palabra.isalnum() and palabra.lower()
    return ' '.join(palabras_limpias)

def limpiar_dataframe(df):
    df['Palabras_Clave'] = df['Palabras_Clave'].apply(limpiar_texto)
    return df

# Función para limpiar HTML
def clean_html(url):
    response = requests.get(url)
    soup = BeautifulSoup(response.content, 'html.parser')
    article_div = soup.find('div', id='article')

    if article_div:
        for script in article_div.find_all('script'):
            script.decompose()
        return article_div.prettify()
    else:
        return None
```
*Ilustración 9. Fragmento del código que realiza la limpieza de los datos previa a la extracción.*

### 4.5 Clasificación de los usuarios por afinidad a tendencias políticas

Una vez tenemos nuestros datos limpios y listos para ser utilizados extraemos un pequeño subconjunto de ellos que anotaremos manualmente. Queremos que el modelo aprenda a predecir la afiliación o afinidad política de un usuario a partir de sus comentarios. Para ello seleccionamos 50 comentarios y de manera manual y según el contexto del comentario les asignamos el signo político más probable.

*Ilustración 10. Anotación manual para el ajuste del modelo mediante few-shot prompting*

A continuación, cargamos en el contexto uno a uno los programas electorales de los diferentes partidos y pedimos al modelo que realice la tarea para el resto de las entradas de nuestro dataset. Siendo posible asignar afinidad política al 93.3% de los comentarios recogidos.

*Ilustración 11. Datos extraídos listos para ser procesados en Pandas.*

5. Resultados

De los datos una vez trabajados podemos obtener algunas conclusiones interesantes. Por ejemplo, en la gráfica 1 podemos observar como se distribuyen los distintos perfiles de votantes que participan comentando en la red social en función de cuales son los aspectos o las áreas de la política francesa que les interesan. Podemos destacar como existe un mucho mayor interés por los asuntos de política interna en los votantes de Reagrupación nacional y como el partido de Emmanuel Macron centra la atención de los asuntos locales.

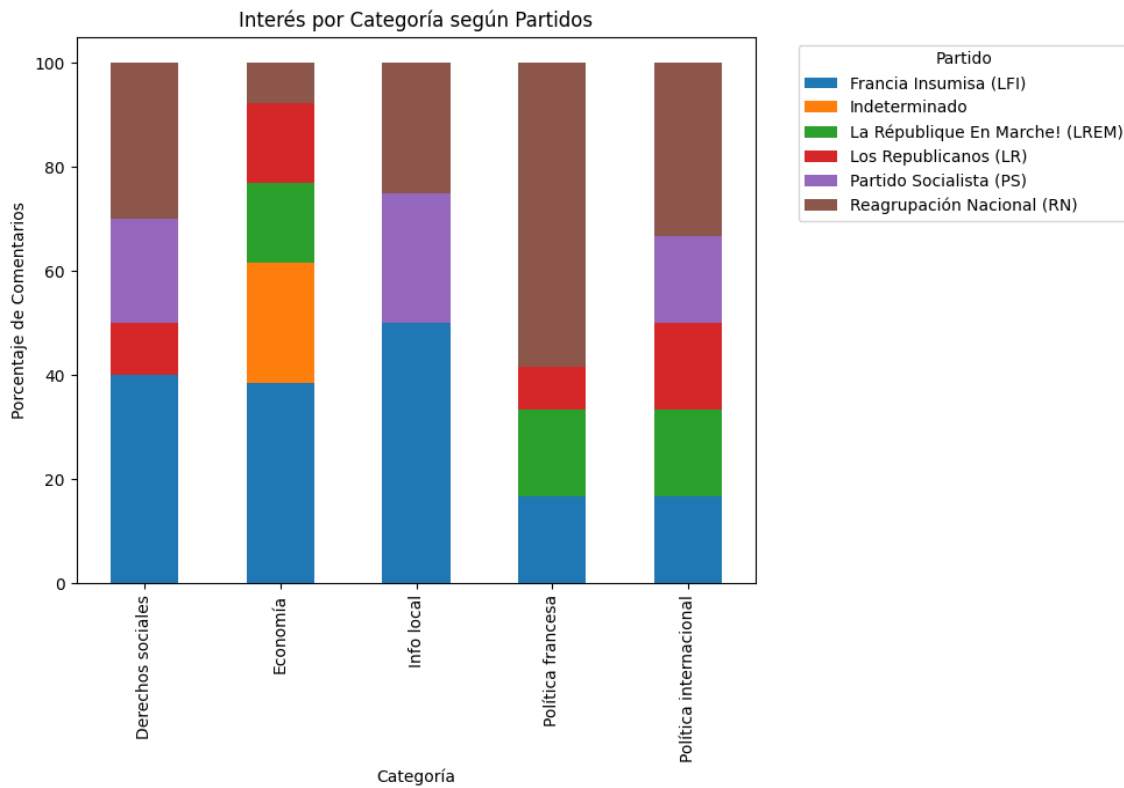

*Gráfica 1. Distribución del interés de los distintos perfiles de votantes por los distintos aspectos de la política francesa.*

La gráfica 2 por otra parte muestra las áreas o los temas que interesan a cada partido político y a los que por tanto se refieren sus propuestas o programas electorales. Aquí puede observarse un fuerte contraste entre los partidos considerados de izquierda y de derecha. Mientras Reagrupación Nacional (RN) muestra un interés predominante en la política francesa y la política internacional, con una atención considerable en los derechos sociales (esto sugiere un enfoque en la soberanía nacional y en las políticas que fortalezcan la identidad y seguridad del país). El partido socialista parece seguir unas líneas más tradicionales confiando en los graneros de voto tradicionales de la izquierda en temas como los derechos sociales, y la justicia social, y dejando espacio a problemas regionales o locales. En fuerte contraste, "En Marche!", el partido de Macron centra su discurso en la política nacional, en aspectos económicos o en política exterior. Francia Insumisa aparece como el partido más transversal lo que le permitiría en principio captar votos a partir de votantes descontentos del resto de partidos o de votantes que provienen del voto de castigo a las actuales políticas gubernamentales.

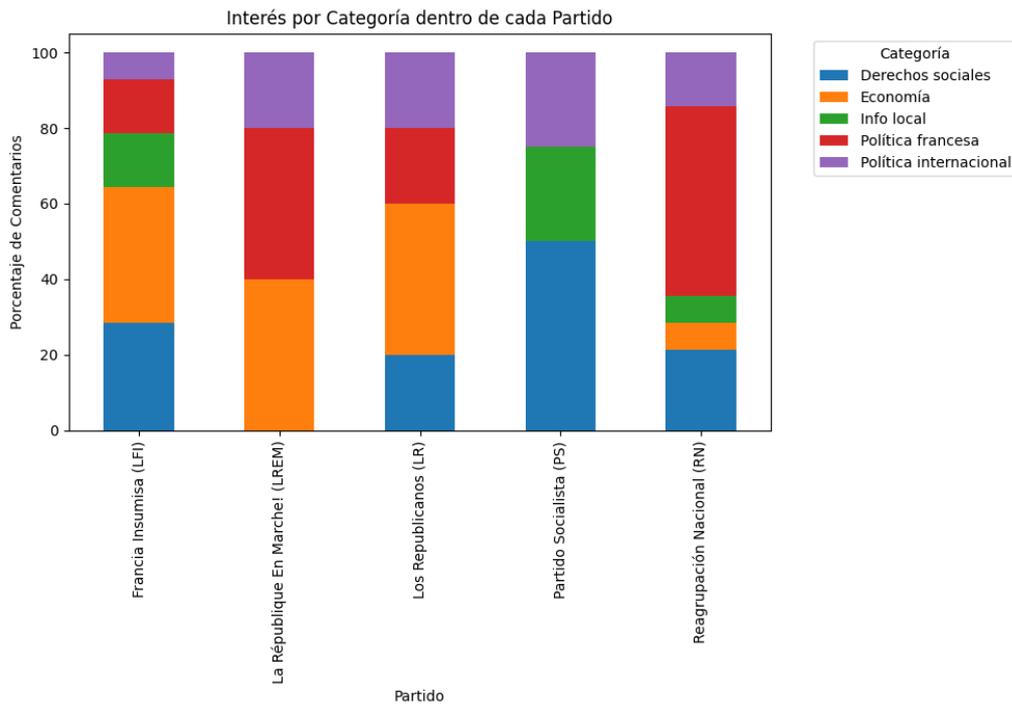

*Gráfica 2. Distribución de temas por partidos según sus programas electorales para las legislativas.*

Cuando realizamos un análisis mas pormenorizado de los términos que aparecen con frecuencia en los discursos de los seguidores de los distintos partidos encontramos que las diferencias las prioridades y los targets de los distintos partidos se reflejan también en vocabulario y la terminología utilizada. Así, los votantes de izquierda tienden a enfocar los distintos temas desde el prisma clásico de la equidad social, y la justicia social. Mientras que los votantes de derecha suelen reconducir los temas bajo el enfoque de la identidad nacional, la soberanía y la seguridad, reflejando una perspectiva más nacionalista.

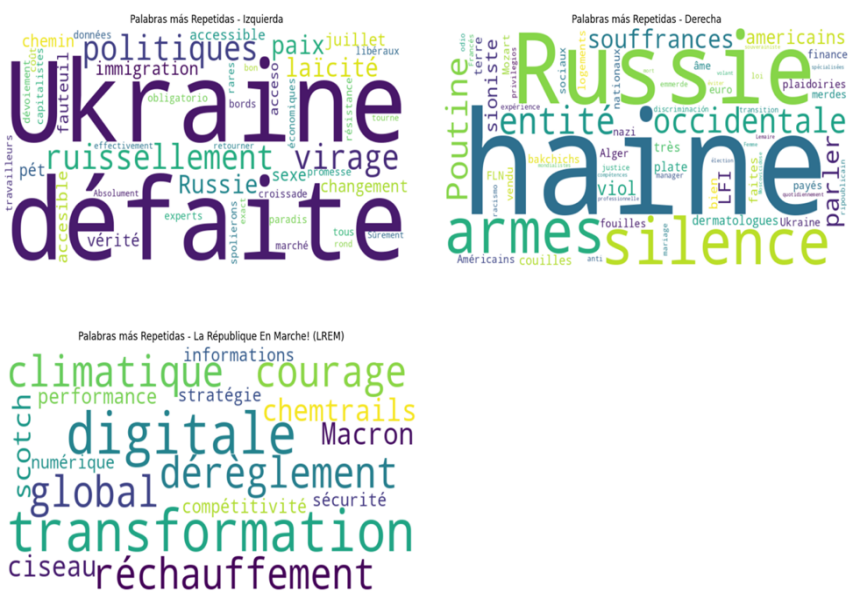

*Ilustración 12. Términos recurrentes en votantes con distintas afinidades.*

Por otro lado, los votantes en los que se enfoca el LREM son personas jóvenes en edad de trabajar y de clase intermedia o acomodada, por lo que sus discursos se centran en la modernización, la competitividad y la sostenibilidad, alineándose con la visión de Macron de una Francia más innovadora y sostenible en un contexto globalizado y europeo. Para este punto hemos creído conveniente agrupar los partidos que concurren juntos en la coalición "Nuevo frente popular" ya que nos interesaba remarcar la diferencia entre los votantes considerados de izquierda, y de derecha.

La lista de términos únicos asociados a los votantes de los distintos perfiles políticos refleja las prioridades, preocupaciones y el lenguaje característico de cada grupo. Los términos de izquierda se centran en temas sociales, y de la denominado justicia social. Palabras como "résistance", "changement", "immigration" y "travailleurs" indican una preocupación por la resistencia al *status quo*, el cambio social, la inmigración y los derechos de los trabajadores, como uno de sus principales nichos de votos. Mientras que términos como "économiques" y "capitalistes" reflejan una crítica al sistema económico actual y al capitalismo. Otros términos característicos para este tipo de votantes incluyen: 'croissade', 'bords', 'bon', 'résistance', 'changement', 'immigration', 'accesible', 'paradis', 'ruissellement', 'retourner', 'économiques', 'capitalistes', 'virage', 'rares', 'libéraux', 'travailleurs', 'rond', 'experts', 'effectivement', 'vérité', 'pét', 'virages', 'acceso', 'politiques', 'laïcité', 'spolierons', 'données', 'Absolument', 'juillet', 'dévoiement', 'sexe', 'chemin', 'défaite', 'paix', 'Ukraine', 'Russie', 'tourne', 'coût', 'promesse', 'exact', 'accessible', 'tous', 'marché', 'obligatorio', 'Sûrement', o 'fauteuil'.

Los términos de derecha se centran en la soberanía, la identidad nacional y la seguridad. Palabras como "souverainiste", "Américains", "communiste", y "racisme" muestran una preocupación por la identidad nacional, la amenaza percibida de influencias extranjeras y una fuerte posición contra el racismo. Términos como "finance", "justice" y "insécurité" indican un enfoque en la estabilidad económica y el orden legal. Además, términos como "violence", "armes", y "sécurité" reflejan una preocupación por la seguridad personal y nacional. Otros términos característicos para este tipo de votantes incluyen: "souverainiste', 'bakchichs', 'Américains', 'âme', 'sociaux', 'sioniste', 'compétences', 'occidentale', 'finance', 'très', 'LFI', 'payés', 'couilles', 'nazi', 'Francés', 'silence', 'professionnelle', 'éviter', 'Alger', 'mariage', 'Lemaire', 'morts', 'plaidoiries', 'faites', 'euro', 'racismo', 'Ukraine', 'justice', 'mondialistes', 'Russie', 'nationaux', 'transition', 'fouilles', 'volant', 'FLN', 'discriminación', 'quotidiennement', 'élection', 'privilèges', 'Femme', 'OTAN', 'expérience', 'armes', 'Moscovicidose', 'Poutine', 'loi', ' gouvernement', 'ripoublicain', 'violence', 'Bruxelles', 'vendu', 'logements', 'americains', 'manager', 'spécialisées', 'terroristes', 'souffrances', 'plate', y 'haine'.

En contraposición a los dos grupos anteriores, los términos mas utilizados en usuarios con afinidad hacia LREM se centran en la modernización, la competitividad y el medio ambiente. Palabras como "compétitivité", "performance", "innovation" y "entrepreneuriat" muestran un enfoque en el crecimiento económico y la innovación. Terminos como "réchauffement", "climatique" y "transition énergétique" reflejan una fuerte preocupación por el cambio climático y la sostenibilidad. Otros términos que encontramos que aparecen con frecuencia incluyen: 'global', 'réchauffement', 'compétitivité', 'ciseau', 'Mondialisation', 'nuclear', 'dérèglement', 'climatique', 'performance', 'chemtrails', 'digitale', 'sécurité', 'transformation', 'informations', 'courage', 'stratégie', 'numérique', 'économie', 'innovation', 'Europe', 'travail', 'éducation', 'transition énergétique', 'entrepreneuriat', 'fiscalité', 'solidarité', 'sécurité sociale', 'souveraineté', y 'environnement'.

6. Conclusiones

Los avances en procesamiento del lenguaje natural, potenciados en los últimos años por el enorme desarrollo que han sufrido lo modelos de embedding, bases de datos vectoriales y los grandes modelos del lenguaje han convertido el análisis de tendencia en redes sociales en una herramienta poderosa para la gestión de la imagen online. Esta evolución tecnológica ha mejorado la capacidad de las herramientas analíticas para detectar patrones, temas emergentes y el impacto de las interacciones en línea sobre la reputación de individuos y organizaciones.

La implementación de modelos avanzados como Word2Vec, GloVe, y especialmente los grandes modelos del lenguaje basados en *Transformer*, ha permitido por primera vez realizar un análisis semántico detallado de grandes volúmenes de datos textuales desestructurados y sin procesar. Estos avances han permitido captar matices y contextos que antes eran inalcanzables, mejorando la precisión del análisis de sentimientos y la clasificación de temas. Además, la capacidad de estos modelos para gestionar datos multilingües y multimodales amplía su aplicabilidad en un entorno global en constante cambio.

En este contexto, el objetivo de nuestro estudio era poner a prueba el poder de los grandes modelos del lenguaje actuales para analizar si los hallazgos sobre la ventaja de Marine Le Pen y su partido, la Agrupación Nacional, en redes sociales masivas como Twitter (X.com) de cara a las próximas elecciones legislativas francesas podrían extrapolarse a otras plataformas digitales menos influenciadas ideológicamente. Para esto, construimos un corpus a partir de uno de los agregadores sociales de noticias más populares con el fin de investigar el grado de aceptación de los mensajes políticos emitidos por diferentes partidos a través de los comentarios y reacciones de los usuarios a las noticias generadas por los medios durante la campaña electoral.

Los resultados del estudio confirmaron que el interés en Reagrupación Nacional sigue siendo alto en comparación con otros partidos. Este resultado coincide con estudios previos que reportan una ventaja significativa de hasta el 400% en redes sociales altamente distribuidas. A pesar de las diversas encuestas preelectorales, el análisis de los datos muestra una clara superioridad en la presencia digital de Marine Le Pen, con un seguimiento más uniforme y distribuido a nivel regional, a diferencia del voto concentrado. en regiones específicas como Île-de-France para La République En Marche! Del actual presidente de la republica Emmanuel Macron.

En términos de interacción y respuesta de los usuarios, los partidos de extrema derecha y de extrema izquierda, representados por la Reagrupación Nacional de Jean-Luc Mélenchon y Francia Insumisa, muestran un aumento notable de comentarios y la difusión de sus mensajes de campaña. Esto sugiere que sus propuestas resuenan más profundamente entre los usuarios de las redes sociales, lo que resulta en una mayor participación digital y una mayor presencia en el discurso público. Por otro lado, partidos con estructuras más tradicionales, como el Partido Socialista y Los Republicanos, parecen tener dificultades para captar atención y generar interacciones en el entorno digital.

7. Código Fuente

El cuaderno y código fuente completo utilizado en este trabajo puede obtenerse desde la url https://colab.research.google.com/drive/1keZ1QvFZVYcaEjsDR-nX43DSy0BhwH9m.

8. Bibliografía